\documentclass[11pt]{revtex4}    
\usepackage{amssymb,epsf}
\usepackage{latexsym}

\begin{document}

\title{Higher dimensional slowly rotating dilaton black holes in AdS spacetime}
\author{A. Sheykhi$^{a,b}$\footnote{sheykhi@mail.uk.ac.ir} and  M.
Allahverdizadeh $^{a}$}
\address{$^a$Department of Physics, Shahid Bahonar University, P.O. Box 76175, Kerman, Iran\\
         $^b$Research Institute for Astronomy and Astrophysics of Maragha (RIAAM), Maragha, Iran}
\begin{abstract}
In this paper, with an appropriate combination of three
Liouville-type dilaton potentials, we obtain the higher
dimensional charged slowly rotating dilaton black hole solution
for asymptotically anti-de Sitter spacetime. The angular momentum
and the gyromagnetic ratio of such a black hole are determined for
the arbitrary values of the dilaton coupling constant. It is shown
that the dilaton field modifies the gyromagnetic ratio of the
rotating dilaton black holes.

\end{abstract}
\maketitle 

\section{Introduction}
Over the past years there has been a growing interest for studying
the rotating black hole solutions in the background of anti-de
Sitter (AdS) spacetimes. This interest is motivated by the
correspondence between the gravitating fields in an AdS spacetime
and conformal field theory on the boundary of the AdS spacetime
\cite{Witt1}. It was argued that the thermodynamics of black holes
in AdS spaces can be identified with that of a certain dual
conformal field theory (CFT) in the high temperature limit
\cite{Witt2}. In the AdS/CFT correspondence, the rotating black
holes in AdS space are dual to certain CFTs in a rotating space
\cite{Haw}, while charged ones are dual to CFTs with chemical
potential \cite{Cham}. The most general higher dimensional
uncharged rotating black holes in AdS space have been recently
found \cite{Haw,Gib}. As far as we know, rotating black holes for
the Maxwell field minimally coupled to Einstein gravity in higher
dimensions, do not exist in a closed form and one has to rely on
perturbative or numerical methods to construct them in the
background of asymptotically flat \cite{kunz1,Aliev2} and AdS
\cite{kunz2} spacetimes. There has also been  recent interest in
constructing the analogous charged rotating solutions in the
framework of gauged supergravity in various dimensions
\cite{Cvetic0,Cvetic1}.

On another front, scalar coupled black hole solutions with
different asymptotic spacetime structure is a subject of interest
for a long time. There has been a renewed interest in such studies
ever since new black hole solutions have been found in the context
of string theory. The low energy effective action of string theory
contains two massless scalars namely dilaton and axion. The
dilaton field couples in a nontrivial way to other fields such as
gauge fields and results into interesting solutions for the
background spacetime \cite{CDB1,CDB2}. These scalar coupled black
hole solutions \cite{CDB1,CDB2}, however, are all asymptotically
flat. It was argued that with the exception of a pure cosmological
constant, no dilaton-de Sitter or anti-de Sitter black hole
solution exists with the presence of only one Liouville-type
dilaton potential \cite{MW}. In the presence of one or two
Liouville-type potentials, black hole spacetimes which are neither
asymptotically flat nor (A)dS have been explored by many authors
(see e.g. \cite{CHM,Cai,Clem,Shey0}). Recently, the ``cosmological
constant term'' in the dilaton gravity has been found by Gao and
Zhang \cite{Gao1,Gao2}. With an appropriate combination of three
Liouville-type dilaton potentials, they obtained the static
dilaton black hole solutions which are asymptotically (A)dS in
four and higher dimensions. The motivations for studying such
dilaton black holes with nonvanishing cosmological constant
originate from supergravity theory. Gauged supergravity theories
in various dimensions are obtained with negative cosmological
constant in a supersymmetric theory. In such a scenario AdS
spacetime constitutes the vacuum state and the black hole solution
in such a spacetime becomes an important area to study
\cite{Witt1}.

In the light of all mentioned above, it becomes obvious that
further study on the rotating black hole solutions in a spacetime
with nonzero cosmological constant in the presence of
dilaton-electromagnetic coupling is of great importance. The
properties of charged rotating dilaton black holes, for an
arbitrary dilaton coupling constant, in the small angular momentum
limit in four \cite{Hor1,Sheykhi2} and higher dimensions have been
studied \cite{Sheykhi3,ShAll}. Recently, in the presence of a
Liouville-type dilaton potential, one of us has constructed a
class of charged slowly rotating dilaton black hole solutions in
arbitrary dimensions \cite{Sheykhi4}. Unfortunately, these
solutions \cite{Sheykhi4} are neither asymptotically flat nor
(A)dS. Besides, they are ill-defined for the string case where
$\alpha=1$. More recently, a class of slowly rotating charged
dilaton black hole solutions in four-dimensional anti-de Sitter
spacetime has been found \cite{Ghosh}. Until now, higher
dimensional charged rotating dilaton black hole solutions for an
arbitrary dilaton-electromagnetic coupling constant in the
background of anti-de Sitter spacetime have not been constructed.
Our aim in this paper is to construct a higher dimensional charged
rotating dilaton black hole solution for asymptotically AdS
spacetime in the small angular momentum limit with an appropriate
combination of three Liouville-type dilaton potentials. We then
determine the angular momentum and the gyromagnetic ratio of such
a black hole for the arbitrary values of the dilaton coupling
constant. We will restrict ourselves to the rotation in one plane,
so our black hole has only one angular momentum parameter.

\section{Field equations and solutions}
We consider the $n$-dimensional $(n\geq4)$ theory in which gravity
is coupled to the dilaton and Maxwell field with an action
\begin{eqnarray}
S &=&-\frac{1}{16\pi }\int_{\mathcal{M}} d^{n}x\sqrt{-g}\left(
R\text{ }-\frac{4}{n-2}\partial_{\mu}\Phi
\partial^{\mu}\Phi-V(\Phi
)-e^{-{4\alpha \Phi}/({n-2})}F_{\mu \nu }F^{\mu \nu }\right)   \nonumber \\
&&-\frac{1}{8\pi }\int_{\partial \mathcal{M}}d^{n-1}x\sqrt{-h
}\Theta (h ),  \label{Act}
\end{eqnarray}
where ${R}$ is the scalar curvature, $\Phi$ is the dilaton field,
$F_{\mu \nu }=\partial _{\mu }A_{\nu }-\partial _{\nu }A_{\mu }$
is the electromagnetic field tensor, and $A_{\mu }$ is the
electromagnetic potential. $\alpha $ is an arbitrary constant
governing the strength of the coupling between the dilaton and the
Maxwell field. The last term in Eq. (\ref{Act}) is the
Gibbons-Hawking surface term. It is required for the variational
principle to be well defined. The factor $\Theta$ represents the
trace of the extrinsic curvature for the boundary ${\partial
\mathcal{M}}$ and $h$ is the induced metric on the boundary. While
$\alpha=0$ corresponds to the usual Einstein-Maxwell-scalar
theory, $\alpha=1$ indicates the dilaton-electromagnetic coupling
that appears in the low energy string action in Einstein's frame.
For an arbitrary value of $\alpha $ in  AdS space the form of the
dilaton potential in arbitrary dimensions is chosen as \cite{Gao2}
\begin{eqnarray}\label{V1}
V(\Phi)&=&\frac{\Lambda}{3(n-3+\alpha^2)^{2}}\left[-\alpha^2(n-2)\left(n^{2}-n\alpha^{2}-6n+\alpha^{2}+9\right)
e^{[{-4(n-3)\Phi}/{(n-2)\alpha}]}\right. \nonumber
\\
&& \left.+(n-2)(n-3)^{2}(n-1-\alpha^{2})
e^{{4\alpha\Phi}/({n-2})}+4\alpha^{2}(n-3)(n-2)^{2}
e^{[{-2\Phi(n-3-\alpha^{2})}/{(n-2)\alpha}]}\right].
\end{eqnarray}
Here $\Lambda $ is the cosmological constant. It is clear the
cosmological constant is coupled to the dilaton in a very
nontrivial way. This type of dilaton potential can be obtained
when a higher dimensional theory is compactified to four
dimensions, including various supergravity models \cite{Gid}. In
the absence of the dilaton field the action (\ref{Act}) reduces to
the action of Einstein-Maxwell gravity with cosmological constant.
Varying the action (\ref{Act}) with respect to the gravitational
field $g_{\mu \nu }$, the dilaton field $\Phi $ and the gauge
field $A_{\mu }$, yields
\begin{equation}
R_{\mu \nu }=\frac{4}{n-2} \left(\partial _{\mu }\Phi
\partial _{\nu }\Phi+\frac{1}{4}g_{\mu \nu }V(\Phi )\right)+2e^{{-4\alpha \Phi}/({n-2})}\left( F_{\mu \eta }F_{\nu }^{\text{
}\eta }-\frac{1}{2(n-2)}g_{\mu \nu }F_{\lambda \eta }F^{\lambda
\eta }\right) ,  \label{FE1}
\end{equation}
\begin{equation}
\nabla ^{2}\Phi =\frac{n-2}{8}\frac{\partial V}{\partial \Phi
}-\frac{\alpha }{2}e^{{-4\alpha \Phi}/({n-2})}F_{\lambda \eta
}F^{\lambda \eta },  \label{FE2}
\end{equation}
\begin{equation}
\partial_{\mu}{\left(\sqrt{-g} e^{{-4\alpha \Phi}/({n-2})}F^{\mu \nu }\right)}=0. \label{FE3}
\end{equation}
We would like to find $n$-dimensional rotating solutions of the
above field equations. For small rotation, we can solve Eqs.
(\ref{FE1})-(\ref{FE3}) to first order in the angular momentum
parameter $a$. Inspection of the $n$-dimensional Kerr solutions
shows that the only term in the metric that changes to the first
order of the angular momentum parameter $a$ is $g_{t\phi}$.
Similarly, the dilaton field does not change to $O(a)$ and
$A_{\phi}$ is the only component of the vector potential that
changes. Therefore, for infinitesimal angular momentum we assume
the metric being of the following form
\begin{eqnarray}\label{metric}
ds^2 &=&-U(r)dt^2+{dr^2\over W(r)}- 2 a f(r)\sin^{2}{\theta}dt
d{\phi}\nonumber \\
 &&+ r^2 R^2(r)\left(d\theta^2 + \sin^2\theta d\phi^2+\cos^2\theta
d\Omega_{n-4}^2\right),
\end{eqnarray}
where $d\Omega^2_{n-4}$ denotes the metric of a unit $(n-4)$-
sphere. The functions $U(r)$, $W(r)$, $R(r)$ and $f(r)$ should be
determined. In the particular case $a=0$, this metric reduces to
the static and spherically symmetric cases. For small $a$, we can
expect to have solutions with $U(r)$ and $W(r)$ still  functions
of $r$ alone. The $t$ component of the Maxwell equations can be
integrated immediately to give
\begin{equation}\label{Ftr}
F_{tr}=\sqrt{\frac{U(r)}{W(r)}}\frac{Q e^{{4\alpha
\Phi}/({n-2})}}{\left( rR\right) ^{n-2}} ,
\end{equation}
where $Q$, an integration constant, is the electric charge of the
black hole. In general, in the presence of rotation, there is also
a vector potential in the form
\begin{equation}\label{Aphi}
 A_{\phi}=- a Q C(r)\sin^2\theta.
\end{equation}
The asymptotically (A)dS static ($a=0$) black hole solution of the
above field equations was found in \cite{Gao2}. Here we are
looking for the asymptotically (A)dS solution in the case
$a\neq0$. Our strategy for obtaining the solution is the
perturbative method suggested by Horne and Horowitz \cite{Hor1}.
Inserting the metric (\ref{metric}), the Maxwell fields
(\ref{Ftr}) and (\ref{Aphi}) into the field equations
(\ref{FE1})-(\ref{FE3}), one can show that the static part of the
metric leads to the following solutions \cite{Gao2}:
\begin{eqnarray}\label{U}
U(r)&=&\left[1-\left(\frac{r_{+}}{r}\right)^{n-3}\right]\left[1-\left(\frac{r_{-}}{r}\right)^{n-3}\right]^{1-\gamma\left(n-3\right)}-\frac{1}{3}\Lambda
r^2\left[1-\left(\frac{r_{-}}{r}\right)^{n-3}\right]^{\gamma} ,
\end{eqnarray}
\begin{eqnarray}\label{W}
W(r)&=&\Bigg{\{}\left[1-\left(\frac{r_{+}}{r}\right)^{n-3}\right]
\left[1-\left(\frac{r_{-}}{r}\right)^{n-3}\right]^{1-\gamma\left(n-3\right)}-\frac{1}{3}\Lambda
r^2\left[1-\left(\frac{r_{-}}{r}\right)^{n-3}\right]^{\gamma}\Bigg
{\}}\nonumber
\\&& \times
\left[1-\left(\frac{r_{-}}{r}\right)^{n-3}\right]^{\gamma(n-4)},
\end{eqnarray}
\begin{eqnarray}\label{Phi}
\Phi(r)&=&\frac{n-2}{4}\sqrt{\gamma(2+3\gamma-n\gamma)}\ln\left[1-\left(\frac{r_{-}}{r}\right)^{n-3}\right],
\end{eqnarray}
\begin{eqnarray}\label{R}
R(r)&=&\left[1-\left(\frac{r_{-}}{r}\right)^{n-3}\right]^{\gamma/2},
\end{eqnarray}
while we obtain the following solution for the rotating part of
the metric
\begin{eqnarray}\label{f0}
f(r)&=&\frac{2\Lambda r^2}{(n-1)(n-2)}\left[1- \left({\frac
{r_{-}}{r}} \right) ^{n-3} \right]^{
\gamma}+\left(n-3\right)\left(\frac{r_{+}}{r}\right)^{n-3}\left[1-\left(\frac{r_{-}}{r}\right)^{n-3}\right]^{({n-3-\alpha^{2}})
/({n-3+\alpha^{2}})}\nonumber\\
&&+\frac{(\alpha^{2}-n+1)(n-3)^{2}}{\alpha^{2}+n-3}r_{-}^{n-3}r^{2}\left[1-\left(\frac{r_{-}}{r}\right)^{n-3}\right]
^{\gamma} \times \int
\left[1-\left(\frac{r_{-}}{r}\right)^{n-3}\right]^{\gamma(2-n)}
\frac{dr}{r^{n}},
\end{eqnarray}
\begin{eqnarray}\label{C}
 C(r)= \frac{1}{r^{n-3}}.
\end{eqnarray}
We can also perform the integration and express the solution in
terms of the hypergeometric function
\begin{eqnarray}\label{f}
f(r)&=&\frac{2\Lambda r^2}{(n-1)(n-2)}\left[1- \left({\frac
{r_{-}}{r}} \right) ^{n-3} \right]^{
\gamma}+\left(n-3\right)\left(\frac{r_{+}}{r}\right)^{n-3}\left[1-\left(\frac{r_{-}}{r}\right)^{n-3}\right]^{(n-3-\alpha^{2})
/(n-3+\alpha^{2})}\nonumber\\
&&+\frac{(\alpha^{2}-n+1)(n-3)^{2}}{(1-n
)(\alpha^{2}+n-3)}(\frac{r_{-}}{r})^{n-3}\left[1-\left(\frac{r_{-}}{r}\right)^{n-3}\right]
^{\gamma}\nonumber\\
&& \times  _{2}F_{1} \left(\left[(n-2)\gamma,\frac
{n-1}{n-3}\right],\left[\frac {2n-4}{n-3}\right],\left({\frac
{r_{-}}{r}}\right) ^{n-3} \right).
\end{eqnarray}
Here $r_+$ and $r_{-}$ are, respectively, the event horizon and
Cauchy horizon of the black hole, and the constant $\gamma$ is
\begin{equation}\label{gamma}
\gamma=\frac{2\alpha^{2}}{(n-3)(n-3+\alpha^{2})}.
\end{equation}
The charge $Q$ is related to $r_+$ and $r_{-}$ by
\begin{equation}\label{Q}
Q^{2}=\frac{(n-2)(n-3)^{2}}{2(n-3+\alpha^{2})}r_{+}^{n-3}r_{-}^{n-3},
\end{equation}
and the physical mass of the black hole is obtained as follows
\cite{Fang}
\begin{equation}\label{mass}
{M}=\frac{\Omega
_{n-2}}{16\pi}\left[(n-2)r^{n-3}_{+}+\frac{n-2-p(n-4)}{p+1}r^{n-3}_{-}\right],
\end{equation}
where we have ignored the term of the order of $a^2$ in the mass
expression for the anti-de Sitter dilatonic black hole \cite
{Ghosh,Aliev3}. Here $\Omega _{n-2}$ denotes the area of the unit
$(n-2)$-sphere and the constant $p$ is
\begin{equation}\label{pp}
{p}=\frac{(2-n)\gamma}{(n-2)\gamma-2}.
\end{equation}
It is apparent that the metric corresponding to
(\ref{U})-(\ref{f}) is asymptotically (A)dS. For $\Lambda=0$, the
above solutions recover our previous results for asymptotically
flat rotating dilaton black holes \cite{ShAll}. In the special
case $n=4$, the static part of our solution reduces to
\begin{equation}\label{U4}
U(r)= W(r)= \left( 1-{\frac {r_{+}}{r}} \right) \left( 1-{\frac
{r_{-}}{r}} \right) ^{
{({1-\alpha}^{2})}/({1+{\alpha}^{2})}}-\frac{1}{3}\Lambda r^2
\left( 1-{\frac {r_{-}}{r}} \right)^{{2\alpha^2}/({1+\alpha^2})},
\end{equation}
\begin{equation}\label{Phi4}
\Phi \left( r \right) =\frac{\alpha}{{\alpha}^{2}+1}\ln \left(1-
{\frac {r_{-}}{r}} \right),
\end{equation}
\begin{equation}\label{R4}
R \left( r \right) = \left( 1-{\frac {r_{-}}{r}} \right) ^{\alpha
^{2}/(1+{\alpha}^{2})},
\end{equation}
while the rotating part reduces to
\begin{eqnarray}\label{fhor}
 f(r)&=&-\left(1-\frac{r_{-}}{r}\right)^{{(1-\alpha^{2})}/(
{1+\alpha^{2}})}\left(1+\frac{(1+\alpha^{2})^{2}r^{2}}{(1-\alpha^{2})(1-3\alpha^{2})
r^{2}_{-}}+\frac{(1+\alpha^{2})r}{(1-\alpha^{2})r_{-}}-\frac{r_{+}}{r}\right)\nonumber\\
&&+\frac{r^{2}(1+\alpha^{2})^{2}}{(1-\alpha^{2})
(1-3\alpha^{2})r^{2}_{-}}\left(1-\frac{r_{-}}{r}\right)^{{2\alpha^{2}}/({1+\alpha^{2}})}+\frac{1}{3}\Lambda
r^2 \left( 1-{\frac {r_{-}}{r}}
\right)^{{2\alpha^2}/({1+\alpha^2})},
\end{eqnarray}
which is the four-dimensional asymptotically (A)dS charged slowly
rotating dilaton black hole solution presented in \cite{Ghosh}.
One may also note that in the absence of a nontrivial dilaton
($\alpha=0=\gamma $), our solutions reduce to
\begin{equation}\label{U0}
U \left( r \right) = W(r)= \left[ 1- \left( {\frac {r_{+}}{r}}
\right) ^{n-3}
 \right]  \left[ 1- \left( {\frac {r_{-}}{r}} \right) ^{n-3}
 \right]-\frac{1}{3}\Lambda r^2,
\end{equation}
\begin{equation}\label{f0}
f \left( r \right)
=(n-3)\left[\frac{r^{n-3}_{-}+r^{n-3}_{+}}{r^{n-3}}-\left(\frac{r_{+}r_{-}}{r^{2}}\right)^{n-3}\right]+{\frac
{2\Lambda\,{r}^{2}}{ \left( n-1 \right)  \left( n-2 \right) }} ,
\end{equation}
which describe the $n$-dimensional charged Kerr-(A)dS black hole
in the limit of slow rotation.

\begin{figure}[tbp]
\epsfxsize=7cm \centerline{\epsffile{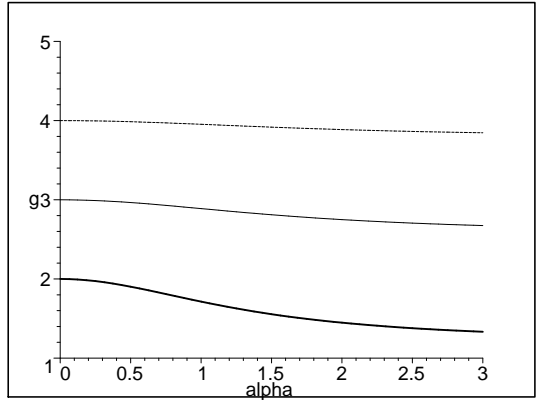}} \caption{The
behavior of the gyromagnetic ratio $g$ versus $\protect\alpha$ in
various dimensions for $r_{-}=1$, $r_{+}=2$. $n=4$ (bold line),
$n=5$ (continuous line), and $n=6$ (dashed line).} \label{figure1}
\end{figure}

Next, we calculate the angular momentum and the gyromagnetic ratio
of these rotating dilaton black holes which appear in the limit of
slow rotation parameter. The angular momentum of the dilaton black
hole can be calculated through the use of the quasilocal formalism
of Brown and York \cite{BY}. According to the quasilocal
formalism, the quantities can be constructed from the information
that exists on the boundary of a gravitating system alone. Such
quasilocal quantities will represent information about the
spacetime contained within the system boundary, just like the
Gauss's law. In our case the finite stress-energy tensor can be
written as
\begin{equation}
T^{ab}=\frac{1}{8\pi }\left(\Theta^{ab}-\Theta h ^{ab}\right) ,
\label{Stres}
\end{equation}
which is obtained by variation of the action (\ref{Act}) with
respect to the boundary metric $h _{ab}$. To compute the
angular momentum of the spacetime, one should choose a spacelike surface $%
\mathcal{B}$ in $\partial \mathcal{M}$ with metric $\sigma _{ij}$,
and write the boundary metric in ADM form
\[
\gamma _{ab}dx^{a}dx^{a}=-N^{2}dt^{2}+\sigma _{ij}\left( d\varphi
^{i}+V^{i}dt\right) \left( d\varphi ^{j}+V^{j}dt\right) ,
\]
where the coordinates $\varphi ^{i}$ are the angular variables
parametrizing the hypersurface of constant $r$ around the origin,
and $N$ and $V^{i}$ are the lapse and shift functions,
respectively. When there is a Killing vector field $\mathcal{\xi
}$ on the boundary, then the quasilocal conserved quantities
associated with the stress tensors of Eq. (\ref{Stres}) can be
written as
\begin{equation}
Q(\mathcal{\xi )}=\int_{\mathcal{B}}d^{n-2}\varphi \sqrt{\sigma }T_{ab}n^{a}%
\mathcal{\xi }^{b},  \label{charge}
\end{equation}
where $\sigma $ is the determinant of the metric $\sigma _{ij}$, $\mathcal{%
\xi }$ and $n^{a}$ are, respectively, the Killing vector field and
the unit normal vector on the boundary $\mathcal{B}$. For
boundaries with rotational ($\varsigma =\partial /\partial \varphi
$) Killing vector field, we can write the corresponding quasilocal
angular momentum  as follows
\begin{eqnarray}
J &=&\int_{\mathcal{B}}d^{n-2}\varphi \sqrt{\sigma
}T_{ab}n^{a}\varsigma ^{b},  \label{Angtot}
\end{eqnarray}
provided the surface $\mathcal{B}$ contains the orbits of
$\varsigma $. Finally, the angular momentum of the black holes can
be calculated by using Eq. (\ref{Angtot}). We find
\begin{equation}
{J}=\frac{a\Omega
_{n-2}}{8\pi}\left(r^{n-3}_{+}+\frac{(n-3)(n-1-\alpha^{2})r^{n-3}_{-}}{(n-3+\alpha^{2})(n-1)}\right).
\label{J}
\end{equation}
For $a=0$, the angular momentum  vanishes, and therefore $a$ is
the rotational parameter of the dilaton black hole. For  $n=4$,
the angular momentum reduces to
\begin{equation}
{J}=\frac{a}{2}\left(r_{+}+\frac{3-\alpha^2}{3(1+\alpha^2)}r_{-}\right),
\label{J4}
\end{equation}
which restores the angular momentum of the four-dimensional Horne
and Horowitz solution \cite{Hor1}. Finally, we calculate the
gyromagnetic ratio of this rotating dilaton black hole. As we
know, the gyromagnetic ratio is an important characteristic of the
Kerr-Newman-AdS black hole. Indeed, one of the remarkable facts
about a Kerr-Newman black hole in asymptotically flat spacetime is
that it can be assigned a gyromagnetic ratio $g = 2$, just as an
electron in the Dirac theory. It should be noted that, unlike four
dimensions, the value of the gyromagnetic ratio is not universal
in higher dimensions \cite{Aliev3}. Besides, scalar fields such as
the dilaton,  modify the value of the gyromagnetic ratio of the
black hole and consequently it does not possess the gyromagnetic
ratio $g = 2$ of the Kerr-Newman black hole \cite{Hor1}. Here, we
wish to calculate the value of the gyromagnetic ratio when the
dilatonic black hole has an asymptotic AdS behavior. The magnetic
dipole moment for this asymptotically AdS slowly rotating dilaton
black hole can be defined as
\begin{equation}\label{mu}
{\mu}=Qa.
\end{equation}
The gyromagnetic ratio is defined as a constant of proportionality
in the equation for the magnetic dipole moment
\begin{equation}\label{mu}
{\mu}=g\frac{QJ}{2M}.
\end{equation}
 Substituting $M$ and $J$ from Eqs. (\ref{mass}) and  (\ref{J}),
 the gyromagnetic ratio $g$ can be obtained as
\begin{equation}\label{g}
g=\frac{(n-1)(n-2)[(n-3+\alpha^2)r^{n-3}_{+}+(n-3-\alpha^2)r^{n-3}_{-}]}{
(n-1)(n-3+\alpha^2)r^{n-3}_{+}+(n-3)(n-1-\alpha^2)r^{n-3}_{-}}.
\end{equation}
One can see that in the linear approximation in the rotation
parameter $a$, the above expression for $g$ turns out to be same
as that found in \cite{ShAll} for the asymptotically flat slowly
rotating dilaton black hole. This means that the dilaton potential
(cosmological constant term) does not change the gyromagnetic
ratio of the rotating (A)dS dilaton black holes, as discussed in
\cite{Ghosh}. However, the dilaton field modifies the value of the
gyromagnetic ratio $g$ through the coupling parameter $\alpha$
which measures the strength of the dilaton-electromagnetic
coupling. This is in agreement with the arguments in \cite{Hor1}.
We have shown the behavior of the gyromagnetic ratio $g$ of the
dilatonic black hole versus $\protect\alpha$ in Fig.
\ref{figure1}. From this figure we find out that the gyromagnetic
ratio decreases with increasing $\alpha$ in any dimension. In the
absence of a nontrivial dilaton $(\alpha=0=\gamma)$, the
gyromagnetic ratio reduces to
\begin{equation}\label{gkerr-newman}
{g}=n-2,
\end{equation}
which is the gyromagnetic ratio of the $n$-dimensional Kerr-Newman
black hole with a single angular momentum in the limit of slow
rotation \cite{Aliev2}. When $n=4$, it reduces to
\begin{equation}\label{gHor}
{g}=2-\frac{4\alpha^{2}r_{-}}{(3-\alpha^{2})r_{-}+3(1+\alpha^{2})r_{+}},
\end{equation}
which is the gyromagnetic ratio of the four-dimensional slowly
rotating dilaton black hole \cite{Hor1,Ghosh}.

\section{Summary and Conclusion}
It is well known that in the presence of one Liouville-type
dilaton potential, no de Sitter or anti-de Sitter dilaton black
hole exists even in the absence of rotation \cite{MW}. In this
paper, with an appropriate combination of three Liouville-type
dilaton potential proposed in \cite{Gao2}, we showed that such
potential leads to higher dimensional slowly rotating charged
dilaton black holes solutions in an anti-de Sitter spacetime. The
presence of such an AdS dilatonic charged rotating black hole is
inevitably associated with an accompanying scalar field with
appropriate Liouville-type potential. Their study therefore may
lead to a better understanding of the origin of the dark matter in
the universe. We started from the nonrotating charged dilaton
black hole solutions in anti-de Sitter spacetime \cite{Gao2} and
then successfully obtained the solution for the rotating charged
dilaton black hole in higher dimensions by introducing a small
angular momentum and solving the equations of motion up to the
linear order of the angular momentum parameter. We discarded any
terms involving $a^2$ or higher powers in $a$ where $a$ is the
rotation parameter. For small rotation, the only term in the
metric which changes is $g_{t\phi}$. The vector potential is
chosen to have a nonradial component $A_{\phi} = -
aQC(r)\sin^2{\theta} $ to represent the magnetic field due to the
rotation of the black hole. As expected, our solution $f(r)$
reduces to the Ghosh and SenGupta solution for $n=4$, while in the
absence of the dilaton field $(\alpha=0=\gamma)$, it reduces to
the $n$-dimensional slowly rotating Kerr-Newman-AdS black hole. We
calculated the angular momentum $J$ and the gyromagnetic ratio $g$
which appear up to the linear order of the angular momentum
parameter $a$. Interestingly enough, we found that the dilaton
field modifies the value of the gyromagnetic ratio $g$ through the
coupling parameter $\alpha$ which measures the strength of the
dilaton-electromagnetic coupling. This is in agreement with the
arguments in \cite{Hor1}.

Finally, we would like to mention that in this paper we only
considered the higher dimensional charged slowly rotating black
hole solutions with a single rotation parameter in the background
of AdS spacetime. In general, in more than three spatial
dimensions, black holes can rotate in different orthogonal planes,
so the general solution has several angular momentum parameters.
Indeed, an $n$-dimensional black hole can have $N = [(n -1)/2]$
independent rotation parameters, associated with $N$ orthogonal
planes of rotation where $[x]$ denotes the integer part of $x$.
The generalization of the present work to the case with more than
one rotation parameter and arbitrary dilaton coupling constant is
now under investigation and will be addressed elsewhere.

\acknowledgments{This work has been supported financially by
Research Institute for Astronomy and Astrophysics of Maragha,
Iran.}


\end{document}